\documentclass[dvips]{acta}
\usepackage{supertabular,lscape,epsfig}
\usepackage{amssymb}
\usepackage{amsmath}
\usepackage{wasysym}
\DeclareSymbolFont{ppa}{OT1}{ppl}{m}{it}
\DeclareMathSymbol{\vv}{\mathalpha}{ppa}{'166}

\newfont{\hb}{rphvb at 10pt}
\newfont{\hbo}{rphvbo at 10pt}
\newfont{\bitt}{rptmbi at 12pt}
\newfont{\bits}{rptmbi at 11pt}

\SetPages{239}{253}

\SetVol{59}{2009}

\begin{document}

\newcommand{\TabCapp}[2]{\begin{center}\parbox[t]{#1}{\centerline{
  \small {\spaceskip 2pt plus 1pt minus 1pt T a b l e}
  \refstepcounter{table}\thetable}
  \vskip2mm
  \centerline{\footnotesize #2}}
  \vskip3mm
\end{center}}

\newcommand{\TTabCap}[3]{\begin{center}\parbox[t]{#1}{\centerline{
  \small {\spaceskip 2pt plus 1pt minus 1pt T a b l e}
  \refstepcounter{table}\thetable}
  \vskip2mm
  \centerline{\footnotesize #2}
  \centerline{\footnotesize #3}}
  \vskip1mm
\end{center}}

\newcommand{\MakeTableSepp}[4]{\begin{table}[p]\TabCapp{#2}{#3}
  \begin{center} \TableFont \begin{tabular}{#1} #4 
  \end{tabular}\end{center}\end{table}}

\newcommand{\MakeTableee}[4]{\begin{table}[htb]\TabCapp{#2}{#3}
  \begin{center} \TableFont \begin{tabular}{#1} #4
  \end{tabular}\end{center}\end{table}}

\newcommand{\MakeTablee}[5]{\begin{table}[htb]\TTabCap{#2}{#3}{#4}
  \begin{center} \TableFont \begin{tabular}{#1} #5 
  \end{tabular}\end{center}\end{table}}

\newfont{\bb}{ptmbi8t at 12pt}
\newfont{\bbb}{cmbxti10}
\newfont{\bbbb}{cmbxti10 at 9pt}
\newcommand{\uprule}{\rule{0pt}{2.5ex}}
\newcommand{\douprule}{\rule[-2ex]{0pt}{4.5ex}}
\newcommand{\dorule}{\rule[-2ex]{0pt}{2ex}}

\hyphenation{OSARGs}
\begin{Titlepage}
\Title{The Optical Gravitational Lensing Experiment.\\
The OGLE-III Catalog of Variable Stars.\\
IV. Long-Period Variables in the Large Magellanic Cloud\footnote{Based on
observations obtained with the 1.3-m Warsaw telescope at the Las Campanas
Observatory of the Carnegie Institution of Washington.}}
\Author{I.~~S~o~s~z~y~ñ~s~k~i$^1$,~~
A.~~U~d~a~l~s~k~i$^1$,~~
M.\,K.~~S~z~y~m~a~ñ~s~k~i$^1$,\\
M.~~K~u~b~i~a~k$^1$,~~
G.~~P~i~e~t~r~z~y~ñ~s~k~i$^{1,2}$,~~
\L.~~W~y~r~z~y~k~o~w~s~k~i$^3$,\\
O.~~S~z~e~w~c~z~y~k$^2$,
~~K.~~U~l~a~c~z~y~k$^1$~~
and~~R.~~P~o~l~e~s~k~i$^1$}
{$^1$Warsaw University Observatory, Al.~Ujazdowskie~4, 00-478~Warszawa, Poland\\
e-mail:
(soszynsk,udalski,msz,mk,pietrzyn,wyrzykow,szewczyk,kulaczyk,rpoleski)
@astrouw.edu.pl\\
$^2$ Universidad de Concepci{\'o}n, Departamento de Fisica, Casilla 160--C,
Concepci{\'o}n, Chile\\
e-mail: szewczyk@astro-udec.cl\\
$^3$ Institute of Astronomy, University of
Cambridge, Madingley Road, Cambridge CB3 0HA, UK\\
e-mail: wyrzykow@ast.cam.ac.uk}
\Received{September 15, 2009}
\end{Titlepage}
\Abstract{The fourth part of the OGLE-III Catalog of Variable Stars
presents 91\,995 long-period variables (LPVs) in the Large Magellanic
Cloud (LMC). This sample consists of 79\,200 OGLE Small Amplitude Red
Giants (OSARGs), 11\,128 semiregular variables (SRVs) and 1667 Mira
stars. The catalog data include basic photometric and astrometric
properties of these stars, long-term multi-epoch {\it VI} photometry and
finding charts.

We describe the methods used for the identification and classification of
LPVs. The distribution of {\it I}-band amplitudes for carbon-rich stars
shows two maxima, corresponding to Miras and SRVs. Such a distinction
between Miras and SRVs is not obvious for oxygen-rich stars. We notice
additional period--luminosity sequence located between Wood's sequences C
and C$'$ and populated by SRVs.}{Stars: AGB and post-AGB -- Stars:
late-type -- Stars: oscillations -- Magellanic Clouds}

\Section{Introduction}
Long-period variables (LPVs) are red giant or supergiant pulsating stars
with periods ranging from about 10 days to a few years. Traditionally, LPVs
are classified into Mira stars, semiregular variables (SRVs) and slow
irregular variables. Miras can be distinguished from SRVs by their regular
light curves and {\it V}-band amplitudes larger than 2.5~mag. Irregular
variables theoretically exhibit no sign of periodicity in their light
curves, but in practice many SRVs with insufficient number of observations
to determine periods are categorized as irregulars. The existence of
strictly non-periodic stars among red giants is a matter of controversy
(\eg Lebzelter and Obbrugger 2009).

Historically, Mira Ceti was the first known non-supernova variable star. Up
to the end of the 19th century Mira-type stars dominated in the catalogs of
variable stars because of their large amplitudes in the visual light. First
LPVs in the Large Magellanic Cloud (LMC) were identified in the first half
of the 20th century, mainly as by-products of searches for Cepheids in this
galaxy (Hoffleit 1937, Shapley and Mohr 1940, McKibben Nail 1952). The
Harvard catalog of variable stars in the LMC by Payne-Gaposchkin (1971)
contained over 100 LPVs, but blue-sensitive surveys allowed to detect only
the brightest red giants.

The first survey dedicated to finding LPVs in the LMC was conducted in the
{\it I}- and {\it V}-passbands by Lloyd Evans (1971). 11 Miras identified
during this search were then observed in the near-infrared (NIR) bands
which allowed Glass and Lloyd Evans (1981) to discover quite narrow
period--luminosity (PL) relation in these wavelengths. Extensive searches
for LPVs in the LMC were undertaken in late 80's by Reid \etal (1988) and
Hughes (1989). These surveys resulted in a discovery of more than 1000
Miras and SRVs in this galaxy.

With the advent of microlensing surveys, like MACHO, OGLE, EROS or MOA
projects, the number of known LPVs was dramatically increased. Wood \etal
(1999) and Wood (2000) used MACHO photometry of about 1500 LPVs in the LMC
to show that these stars follow five distinct PL sequences (labeled
A--E). This result was confirmed by many other studies (\eg Noda \etal
2002, Lebzelter \etal 2002, Cioni \etal 2003). The largest catalog of LPVs
in the LMC available to date was compiled by Fraser \etal (2008) on the
basis of the MACHO photometry. This sample comprises 56\,453 stars of
which 32\,899 objects are associated with one of the Woods' PL sequences.

The long-term photometry collected during the OGLE survey allowed to
greatly enhance our knowledge about LPVs. The OGLE-II light curves combined
with the NIR photometry originated in various sources were studied by Kiss
and Bedding (2003, 2004), Ita \etal (2004ab), Groenewegen (2004) and Wray
\etal (2004). We extended the time span of the observations by
supplementing the OGLE-II photometry with the OGLE-III data and analyzed
various aspects of the red giant variability: OGLE Small Amplitude Red
Giants (OSARGs, Soszyñski \etal 2004a), ellipsoidal and eclipsing binaries
that follow sequence E in the PL diagrams (Soszyñski \etal 2004b), Miras
and SRVs (Soszyñski \etal 2005) and still unexplained Long Secondary Period
(LSP) phenomenon that is responsible for sequence~D (Soszyñski 2007). In
the paper by Soszyñski \etal (2007) we extended to 14 the number of
individual sequences in the period \vs Wesenheit index plane.

In this paper we describe the catalog of 91\,995 pulsating red giant stars
identified in the 40 square degrees of the LMC observed regularly between
2001 and 2009 in the course of the third stage of the OGLE project
(OGLE-III). This is the fourth part of the OGLE-III Catalog of Variable
Stars (OIII-CVS). In the previous parts we presented other types of
pulsating stars in the LMC: classical Cepheids (Soszyñski \etal 2008a),
type~II and anomalous Cepheids (Soszyñski \etal 2008b) and RR~Lyr stars
(Soszyñski \etal 2009).

The paper is structured as follows. In Section~2 we describe the OGLE
photometric data. Section~3 gives the details about the process of LPVs
selection and classification. In Section~4 we describe the catalog itself,
and in Section~5 we compare our sample with the catalog of LPVs by Fraser
\etal (2008). In Section~6 we discuss and summarize our results.

\Section{Observations and Data Reduction}
The long time span of observations is crucial in the detection and analysis
of LPVs. The OGLE-III photometry was accumulated over the interval of
almost 8 years, from July 2001 to May 2009. For stars in the central 4.5
square degrees of the LMC the OGLE-II photometry is available, which
increases the time baseline to 13 observing seasons.

All the {\it VI} standard photometry was obtained with the 1.3 meter Warsaw
telescope located at Las Campanas Observatory in Chile. The observatory is
operated by the Carnegie Institution of Washington. During the OGLE-III
phase the telescope was equipped with the eight-chip CCD mosaic camera of
the total resolution $8192\times8192$ pixels and the field of view of about
$35\times35.5$ arcmin. For details of the instrumental setup we refer to
Udalski (2003).

The 116 OGLE-III fields cover almost 40 square degrees in the LMC. The
majority of the observations were taken in the {\it I}-band, typically
about 500 points per star. For stars with the OGLE-II observations the
number of points reaches 1000. About 40--60 observations per star
($\approx100$ with the OGLE-II data) were secured in the {\it V}-band. The
OGLE data reduction pipeline is based on the Difference Image Analysis
technique (DIA; Wo¼niak 2000, Udalski 2003) which produces much better
photometry than standard profile-fitting methods. However, the OGLE-III
frames were also reduced using program {\sc DoPhot} (Schechter \etal 1993)
and independent PSF photometry and astrometry were obtained for all the
stars. In this catalog we use the {\sc DoPhot} photometry for about 500
bright stars that are saturated in the DIA reference frames. These objects
are flagged in the remarks of the catalog. Full description of the
reduction techniques, photometric calibration and astrometric
transformations can be found in Udalski \etal (2008a).

To select and classify LPVs in the LMC we used single-epoch near-infrared
(NIR) photometry originated in two sources: 2MASS All-Sky Catalog of Point
Sources (Cutri \etal 2003) and the IRSF Magellanic Clouds Point Source
Catalog (Kato \etal 2007). Since LPVs are sometimes heavily reddened by
circumstellar matter, we derived for each star the extinction-free
Wesenheit indices using the NIR and visual magnitudes:
$$W_{JK}=K_s-0.686(J-K_s)$$
$$W_I=I-1.55(V-I)$$
Note that in the above equation $I$ and $V$ are not just the mean
magnitudes calculated for our light curves. Because, in general, time
coverage of the OGLE observations is different in $I$ and $V$ domains and
LPVs sometimes change their mean luminosity in time, we decided to derive
individual $W_I$ values for each $V$ point (approximating {\it I}-band
magnitude for a given epoch using three closest points in the light
curve). Then, the series of $W_I$ points were treated as a normal light
curves and the intensity mean values of the Wesenheit index were derived by
a fitting of a Fourier series.

\Section{Identification and Classification of LPVs}
We began the selection of LPVs in the LMC with the period search for all
stars brighter than $I=18$~mag in the OGLE-III database and $K_s=14$~mag in
the 2MASS catalog. In the further analysis we also checked the
periodicities of fainter stars, but with large values of the $(J-K_s)$
index. We searched the frequency space 0--0.5~d$^{-1}$ with a resolution
$10^{-6}$~d$^{-1}$ using the {\sc Fnpeaks} code (Z.~Ko{\l}aczkowski, private
communication). For each star the procedure was as follows. After finding
the primary period the third order Fourier series was fitted to the light
curve and this function was subtracted from the data. Then, this step was
repeated on the residual data, until we found 15 periods per star.

\begin{figure}[p]
\centerline{\includegraphics[width=13.5cm]{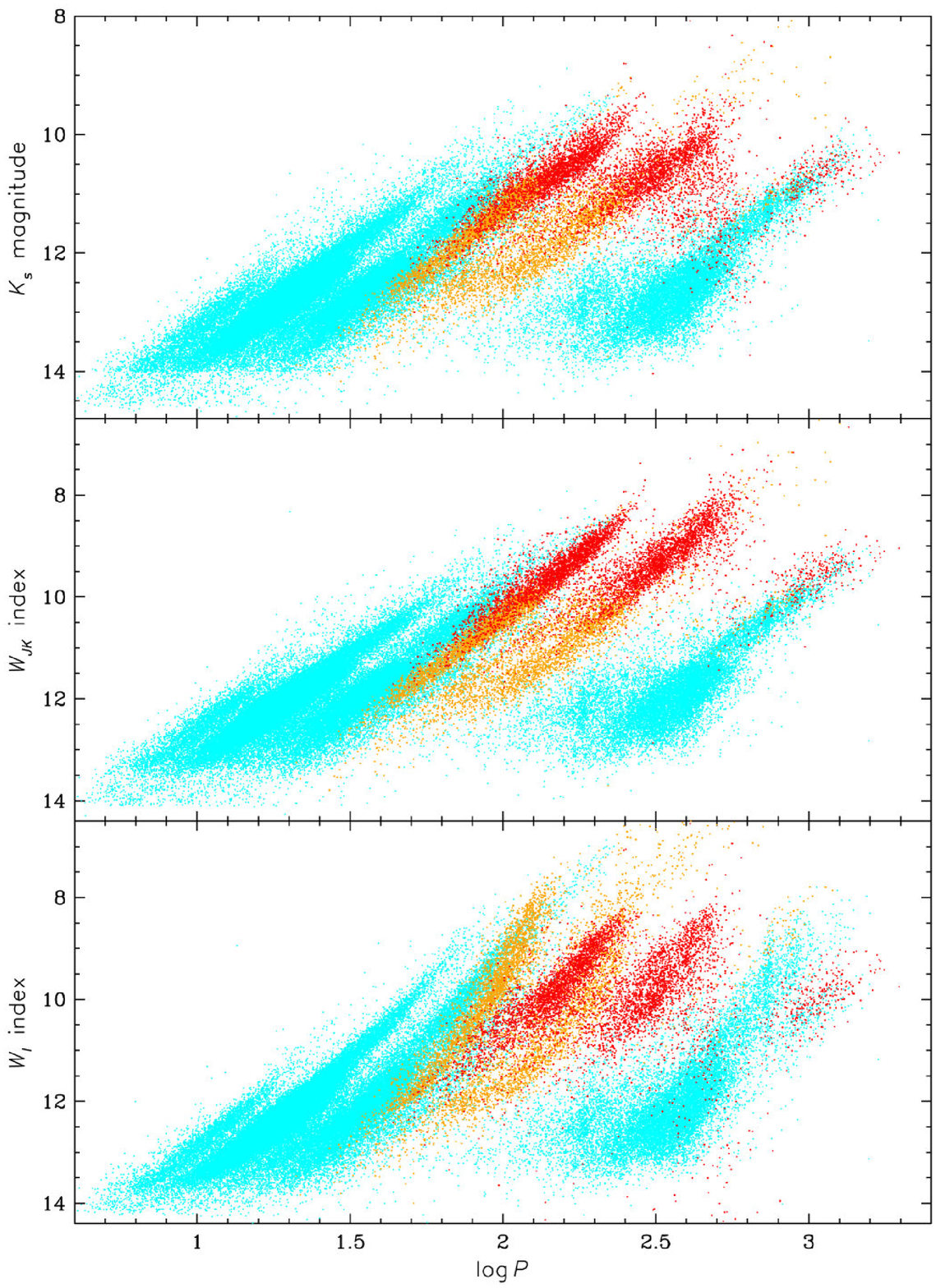}}
\FigCap{Period--luminosity diagrams for LPVs in the LMC. 
{\it Upper, middle} and {\it bottom panels} show $\log{P}$--$K_s$,
$\log{P}$--$W_{JK}$ and $\log{P}$--$W_I$ diagrams, respectively. Cyan
points mark OSARG variables, orange points indicate O-rich Miras and SRVs,
and red points show positions of C-rich Miras and SRVs. Each star is
represented by one (the primary) period.}
\end{figure}
The stars were plotted in the PL diagrams (Fig.~1) and the well-known
series of sequences (Soszyñski \etal 2007) appeared. The location in the
$\log{P}$--$W_{JK}$ diagram was the primary criterion used to select and
classify LPVs. The {\it I}-band light curves of stars with the primary
periods in sequences C and C$'$ were inspected by eye. In some cases the
primary periods were judged to be spurious and new periods were
found. Stars which followed other PL sequences or had primary period
outside any of the ridge were visually inspected only if their amplitude of
variability was larger than about 0.1~mag. Again, in some cases we decided
to correct the primary periods.

Classification of LPVs is a complex issue, because of the complicated
nature of their light curves which often show multiperiodicity, irregular
variations, changes of the mean magnitudes or modulations of periods,
phases and amplitudes. Moreover, there is a contribution of young stellar
objects (YSO), quasars in the background, and Galactic red dwarfs in the
foreground of the LMC among red variable objects.

To filter out these undesirable objects we used the recently published
catalog of YSO in the LMC (Gruendl and Chu 2009) and the list of candidates
for quasars behind the LMC (Koz³owski and Kochanek 2009). We
cross-identified objects from these catalogs with our preliminary list of
red variable stars in the LMC and inspected their light curves. With the
exception of a few cases which were judged to be evident LPVs, we removed
these objects from our catalog. To exclude Galactic red stars from the
sample we checked the proper motions of all stars in our preliminary
list. We used the {\sc DoPhot} astrometric measurements obtained
independently for each observation. Stars with measurable proper motions
(down to about 2.5~mas/year) were recognized as nearby red dwarfs and
removed from the sample.

The stars in the catalog are divided into three main classes: OSARGs, SRVs
and Miras. With the exception of several R~CrB stars (which will be
published in another part of the OIII-CVS), we did not notice red giants
with no sign of periodicity, \ie irregular variables. Also, red giants
showing eclipsing or ellipsoidal variations (sequence~E stars) are not
included in this part of the OIII-CVS, unless they exhibit pulsational
variability superimposed on the primary periods.

OSARGs constitute the most numerous class of LPVs, and this is probably the
most numerous class of variable stars in the LMC at all (at least in the
range of luminosities and amplitudes available for the Warsaw
telescope). Note that in the traditional classification scheme (Kholopov
\etal 1985) there is no such type of LPVs as OSARGs. Most of the OSARGs
from our catalog (if they were detected at all despite their small
amplitudes) would be categorized as SRVs. However, Soszyñski \etal (2004a)
noticed that OSARG variables characterize with the unique features,
different than for ``classical'' SRVs. For example, OSARGs obey different
PL relations than SRVs and Miras. Also the Petersen diagrams show
completely different patterns for OSARGs and SRVs.

Although OSARGs and SRVs seem to belong to two different types of variable
stars, their distinction is not an easy task, because both groups partly
overlap in all diagrams that can be plotted using the OGLE data. To
separate OSARGs and SRVs we utilized the algorithm described by
Soszyñski \etal (2007). In brief, this algorithm uses characteristic period
ratios observed for OSARG variables, and their positions in the
period \vs Wesenheit index diagrams. The larger number of periods in a given
object which fall within the OSARG PL sequences, the higher probability
that the star belong to this class of variable stars.

The majority of OSARGs in our catalog were selected with this
method. Sometimes, our algorithm failed, for example for blended stars
(which do not lie exactly on the PL relations), or for stars with no
measurements in one of the passband which are used to construct Wesenheit
indices, or just for stars with insufficient number of detectable periods
that place a star in various PL sequences. These objects were treated
individually. We categorized them as OSARGs if their parameters located
them in the region occupied by OSARGs in the period--luminosity,
period--amplitude and color--magnitude diagrams.

In total, 79\,200 stars were classified as OSARGs. Note, that we inspected
by eye only a small fraction of their light curves, so the contribution of
spurious detections is possible. The primary periods provided in our
catalog are not always the pulsation periods which locate the stars on one
of the OSARG sequences a$_1$--a$_4$ or b$_1$--b$_3$ (Soszyñski \etal
2007). In many cases the primary periods belong to sequence~D (LSP) or E
(eclipsing or ellipsoidal variables).

OSARG variables were divided into two groups: first-ascent red giant branch
stars (RGB) and asymptotic giant branch stars (AGB). Of course, all the
stars brighter than the tip of the RGB ($K_s=12.05$~mag) are AGB
variables. Below the tip we used the criterion discovered by Soszyñski
\etal (2004a). If any of the period falls on the shortest-period sequence
a$_4$, the star was categorized as an AGB variable. The remaining OSARGs
are classified as RGB stars, however one should remember that there might
be a subset of AGB stars that have no periods in sequence a$_4$ among these
objects.

The remaining LPVs were classified as Mira stars and SRVs. Most of them lie
on sequence C or C$'$, and very often on both sequences. However, there is
no doubt that another, dimmer PL ridge is located between sequences C and
C$'$ (see Section~6). There is also a number of SRVs with the primary
periods on sequence~D.

\begin{figure}[htb]
\centerline{\includegraphics[width=11.3cm, bb=65 480 555 745]{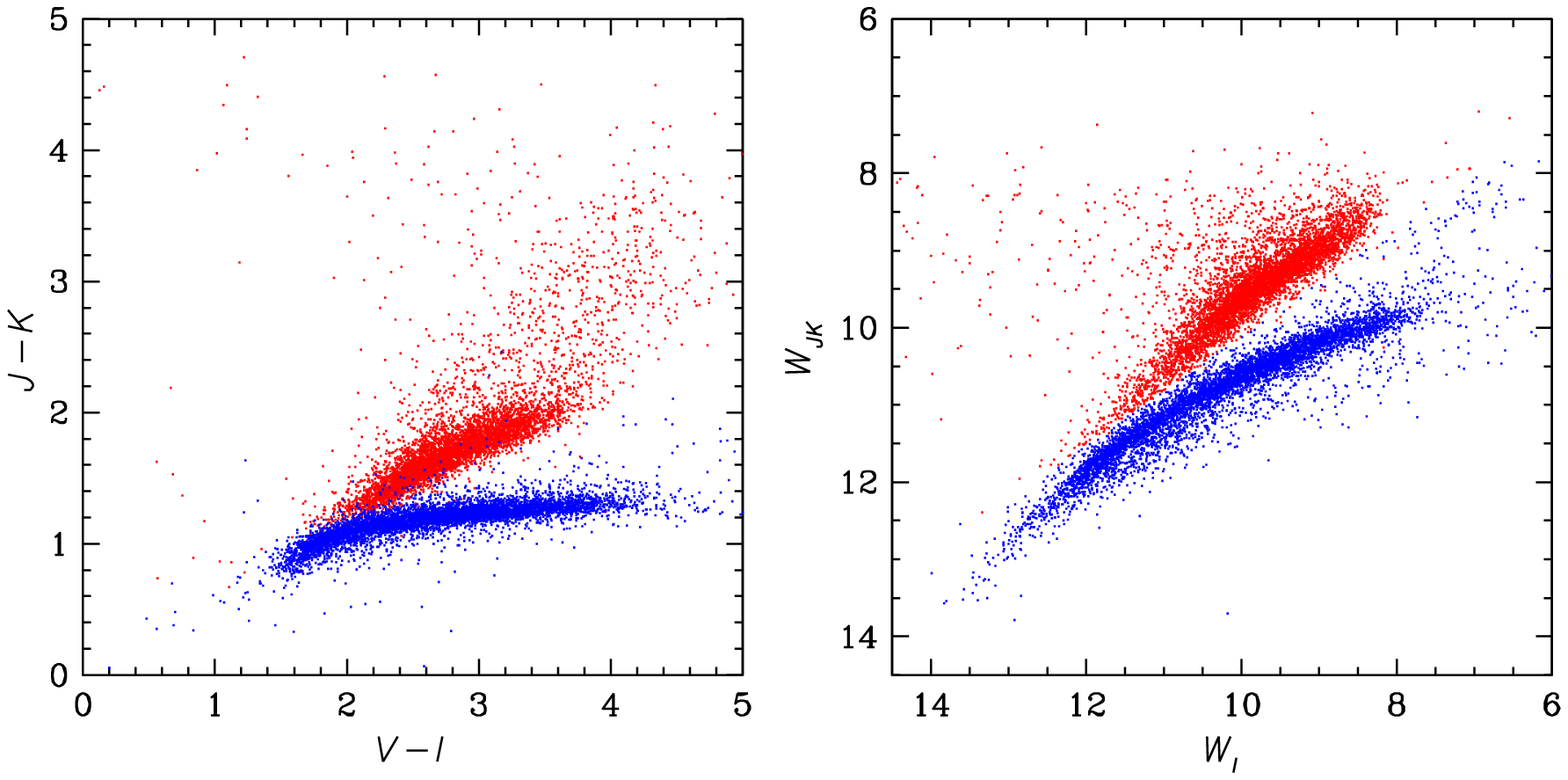}}
\FigCap{Color--color ({\it left panel}) and $W_I$--$W_{JK}$ ({\it right panel})
diagrams for Miras and SRVs in the LMC. Blue points show O-rich, while red
points represent C-rich stars.}
\end{figure}
We divided Miras and SRVs into oxygen-rich (O-rich) and carbon-rich
(C-rich) stars. Soszyñski \etal (2005) showed that both classes can be
distinguished on the period \vs $W_I$ diagram. However, the additional ridge
between sequences C and C$'$ complicates this pattern, so we used another
method to discriminate between O-rich and C-rich stars. Both classes are
well separated in the $(V-I)$ \vs $(J-K)$ color--color diagram (Fig.~2, left
panel) and even better distinguished in the $W_I$--$W_{JK}$ diagram
(Fig.~2, right panel). The later plane was used to separate O-rich and
C-rich stars. Note that our classification is based on the photometric
measurements only and should be confirmed spectroscopically. Moreover,
there is a region in the diagrams where both populations overlap, so our
classification may be wrong for stars in these surroundings. It is possible
that this region is occupied by stars in an intermediate stage between
O-rich and C-rich giants, \ie S-type stars.

Formally, Miras and SRVs can be distinguished by their amplitudes. In the
General Catalogue of Variable Stars (GCVS; Kholopov \etal 1985) the
limiting amplitude is defined at 2.5~mag in the {\it V}-band, however even
within the GCVS this rule is not strictly obeyed. One should also remember
that Miras and some of the SRVs follow the same PL relation, \ie
sequence~C, so both type of stars must be closely related to each other.

Since the {\it V}-band measurements are sparse in the OGLE database, we
decided to use {\it I}-band amplitudes to discriminate between Miras and
SRVs. Before we derived the amplitudes the light curves had been detrended
by fitting and subtracting cubic splines. It was necessary especially for
C-rich stars, because many of them exhibit strong irregular variations of
their mean brightness. The third-order Fourier series were fitted to the
light curves. The amplitudes $A_I$ were defined as the differences between
maximum and minimum values of these functions.

\begin{figure}[htb]
\centerline{\includegraphics[width=11.3cm, bb=65 400 555 745]{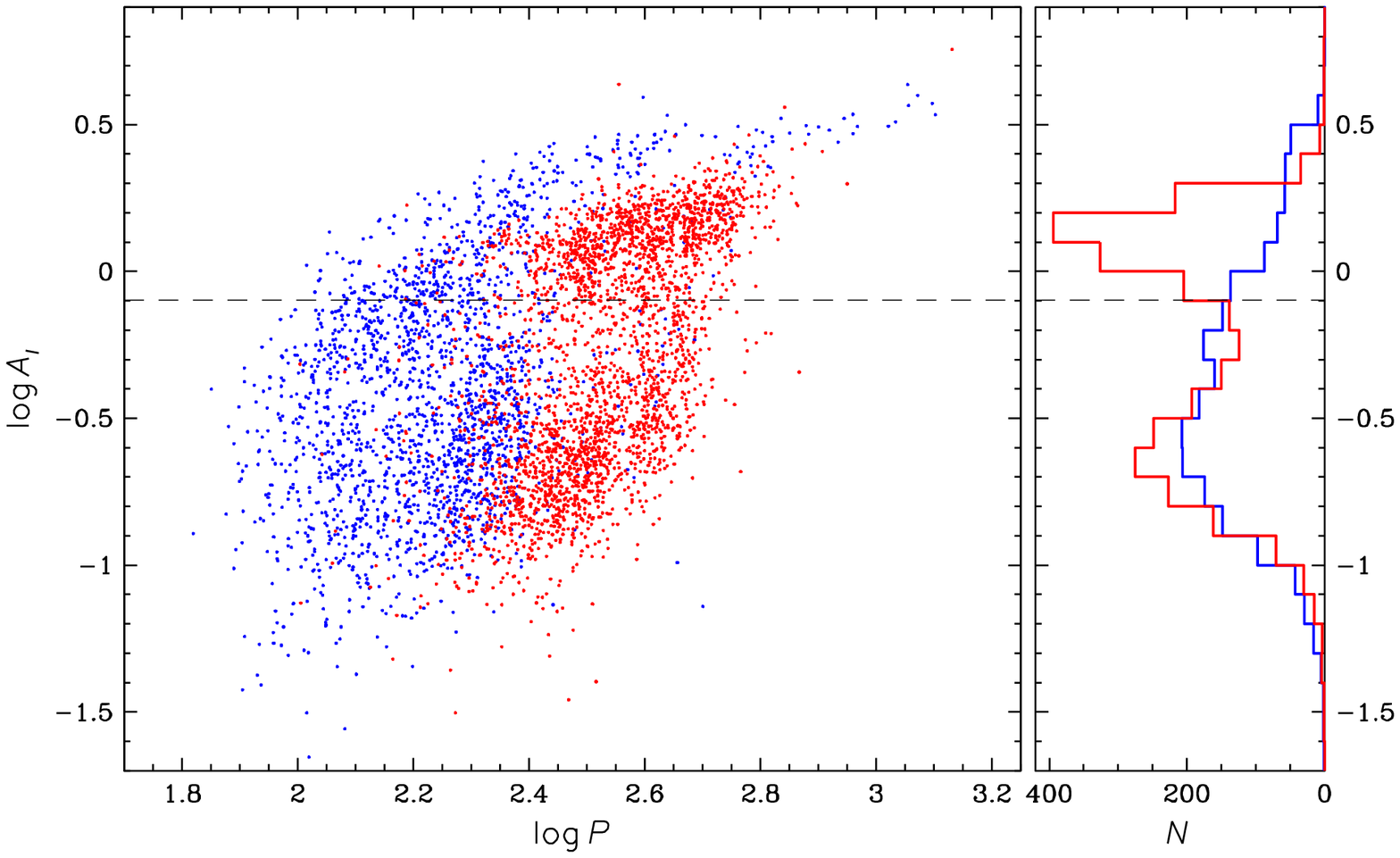}}
\FigCap{Period--amplitude diagram ({\it left panel}) for the sequence C stars
(Miras and SRVs). Blue points show O-rich, while red points represent
C-rich stars. The {\it right panel} presents the distribution of amplitudes
for O-rich (blue line) and C-rich (red line) stars. Black dashed line marks
the limiting amplitude ($A_I=0.8$~mag) used to separate Miras and SRVs in
this paper.}
\end{figure}
Fig.~3 shows the period--amplitude diagram for stars classified as SRVs and
Miras and located in sequence~C. In the right panel we show histograms of
$\log{A_I}$ separately for O-rich and C-rich stars. C-rich stars evidently
have bimodal distribution of amplitudes with local maxima at $A_I=0.25$~mag
and $A_I=1.3$~mag. This feature clearly defines the transition between
Miras and SRV. For O-rich stars such a bimodality of the amplitude
distribution is invisible or barely visible. After all, we decided to use
the same limiting amplitude $A_I=0.8$~mag for both spectral types to
distinguish between Miras and SRVs. The dashed line shows this boundary
value in Fig.~3.

\Section{The Catalog}
The OGLE-III Catalog of LPVs in the LMC comprises 91\,995 stars, of which
79\,200 are classified as OSARG variables, 11\,128 as SRVs and 1667 as Mira
stars. OSARG variables are divided into 45\,528 RGB and 33\,672 AGB
stars. Miras and SRVs consist of 6445 O-rich and 6350 C-rich giants. In
Fig.~4 we present spatial distribution of our samples, separately for OSARG
variables, O-rich and C-rich Miras and SRVs. It is clearly seen that C-rich
stars are more concentrated toward the LMC bar than O-rich variables.
\begin{figure}[t]
\centerline{\includegraphics[width=12.7cm]{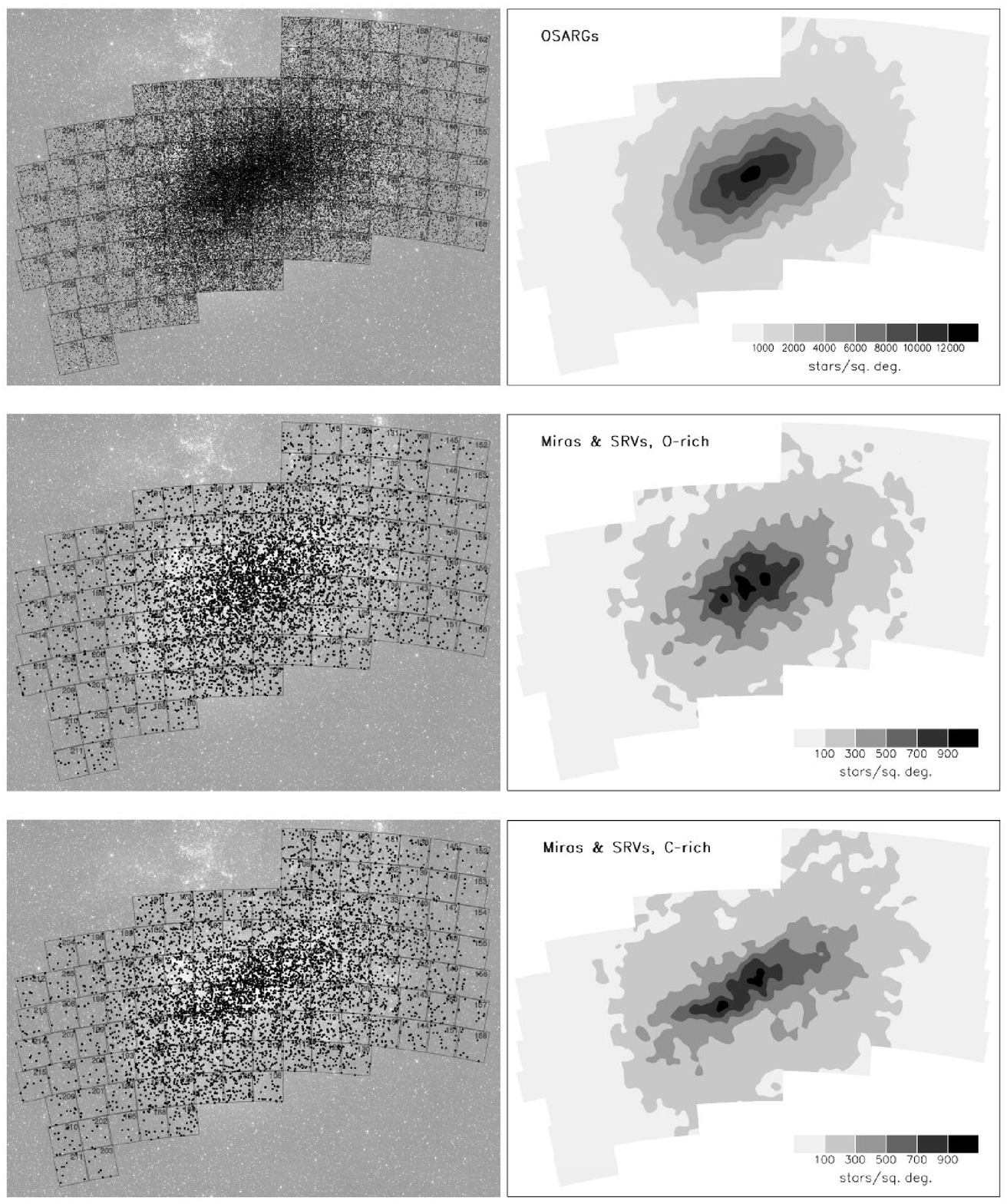}}
\vspace*{2mm}
\FigCap{Spatial distribution of LPVs in the LMC. {\it Left panels} 
show positions of stars overplotted on the LMC image originated from the
ASAS sky survey (Pojmañski 1997). Right panels present surface density maps
obtained by smoothing of the distributions with the Gaussian
filter. {\it Upper panels} show OSARG variables, {\it middle panels} --
O-rich SRVs and Miras and {\it bottom panels} -- O-rich SRVs and Miras.}
\end{figure}

The catalog data are available on-line from the OGLE Internet Archive
through the WWW interface or {\it via} anonymous FTP site:
\begin{center}
{\it http://ogle.astrouw.edu.pl/} \\ 
{\it ftp://ftp.astrouw.edu.pl/ogle/ogle3/OIII-CVS/lmc/lpv/}\\
\end{center}

In the FTP site the full list of LPVs is given in the file {\sf
ident.dat}. Stars are listed in order of increasing right ascension and
designated with symbols OGLE-LMC-LPV-NNNNN, where NNNNN is a five digit
consecutive number. The file {\sf ident.dat} contains the following
information about each star: the object designation, a cross-identification
with the OGLE-III photometric maps of the LMC (Udalski \etal 2008b; lack
of the identification means that the star is saturated in the DIA
database and we publish the DoPhot photometry or the OGLE-II photometry
only), classification (Mira, SRV, OSARG),
evolutionary status (RGB, AGB), spectral type (O-rich, C-rich), equinox
J2000.0 right ascension and declination, cross-identifications with the
OGLE-II database (Szymañski 2005), with the MACHO catalog of LPVs in the
LMC (Fraser \etal 2008), with the extragalactic part of the GCVS
(Artyukhina \etal 1995).

Physical parameters of the stars -- mean $I$ and $V$ magnitudes, periods,
and amplitudes -- are provided in the files {\sf OSARGs.dat}, {\sf
SRVs.dat}, and {\sf Miras.dat}. The multi-epoch {\it I}- and {\it V}-band
photometry is written in separate files in the subdirectory {\sf
phot/}. The subdirectory {\sf fcharts/} contains finding charts for all
objects. These are the $60\arcs\times60\arcs$ subframes of the {\it I}-band
DIA reference images, oriented with N up, and E to the left. The file {\sf
remarks.txt} contains additional information about some objects. Here we
flag the stars with the LSP (periods in sequence D) or
ellipsoidal/eclipsing variations (sequence E), but this is not a full list
of such objects. Our selection includes only about 7000 of the
largest-amplitude sequence D variable stars and about 2000
eclipsing/ellipsoidal red giants.

For each star we provide three periods. Only for Miras and SRVs and only
the primary periods were visually checked and corrected, if
necessary. Thus, these primary periods may not correspond to the highest
peak in the power spectrum. For example, Fourier analysis of a light curve
with strong irregular component sometimes gives dominant period than cannot
be associated with any real period. In such case we replaced the primary
period with one selected from the secondaries. The second and the third
periods provided in the catalog tables were derived fully automatically
after prewhitening the light curves with the first periods. These secondary
periods may not by associated with any real periodicity. For the complex
frequency analyzes of our sample of LPVs we recommend to perform
independent period search using the photometry attached to this catalog.

\Section{Completeness of the Catalog}
The completeness of our sample strongly depends on the amplitudes of
variations. The catalog is nearly complete for Miras (with the exception of
objects that are too bright and saturate in our data, or too faint -- below
the detection limit). On the other side are the faintest OSARG variables
which simultaneously have the smallest amplitudes, on the detection limit
of the OGLE photometry. It cannot be ruled out that despite of our cleaning
procedure there is still a small contribution of different types of red
variable objects: foreground stars, YSO, quasars, etc. in our sample of
LPVs.

To test the completeness of the catalog we compared it with the MACHO
catalog of LPV in the LMC recently published by Fraser \etal (2008).
48\,141 of stars in this catalog could be found in the OGLE-III
fields, but only 28\,257 of these variable stars are assigned by
Fraser \etal (2008) to one of the sequences in the PL space. The remaining
objects are classified as ``one-year artifacts'', are outside the
boundaries of any PL sequence, or have no classification at all. These
stars are not firm LPVs, so we removed them from the sample which were
matched with our catalog.

We cross-identified the OGLE and MACHO lists and found no counterparts for
1208 MACHO variables. We carefully checked all these stars which are not
present in our sample. 716 of these objects are sequence~E stars
(ellipsoidal and eclipsing binaries) which will be published in the
future. 98 of missing objects occurred to be Cepheids, RR Lyr stars, young
stellar objects, quasars, nearby red dwarfs or other types of variable
objects. From the remaining 394 stars (1.4\% of the MACHO sample) 134
objects had extremely low amplitudes (we included some of them in our
catalog, for others we did not detect any significant variability), 81 were
affected by a small number of observations due to their location close to
the edge of the OGLE field and 179 were saturated in our frames.

\Section{Discussion}
The catalog of LPVs in the LMC is the largest part of the OIII-CVS
published to date. It contains the most heterogeneous sample of variable
stars. The catalog includes stars as bright as $I=12$~mag and as faint as
$I=20$~mag. The amplitudes of variations cover a range from millimagnitudes
to several magnitudes. Periods range from a few days to years.

OSARGs which constitute the vast majority of our catalog are one of the
less studied variable stars. In the paper by Soszyñski \etal (2007) we
suggested that the mechanism responsible for the OSARG variability may be
a stochastic (solar-like) excitation. It
is still an open question if there exist stars which exhibit features of
both classes -- OSARGs and ``classical'' SRVs (or Miras). Our
classification procedure is not perfect and cannot definitely answer this
question. Most of the OSARG variables are O-rich stars. Fig.~4 shows that
the spatial distribution of OSARGs is similar to the distribution of O-rich
SRVs and Miras. However, about 3300 OSARGs are located in the region of
C-rich stars in the $W_I$--$W_{JK}$ diagram (Fig.~2). Some of these stars
may have spuriously measured magnitudes, but there is no doubt that true
C-rich OSARGs also exist. We found even a few candidates for R~Crb stars
among C-rich OSARGs.

\begin{figure}[b]
\centerline{\includegraphics[width=11.3cm, bb=65 235 535 745]{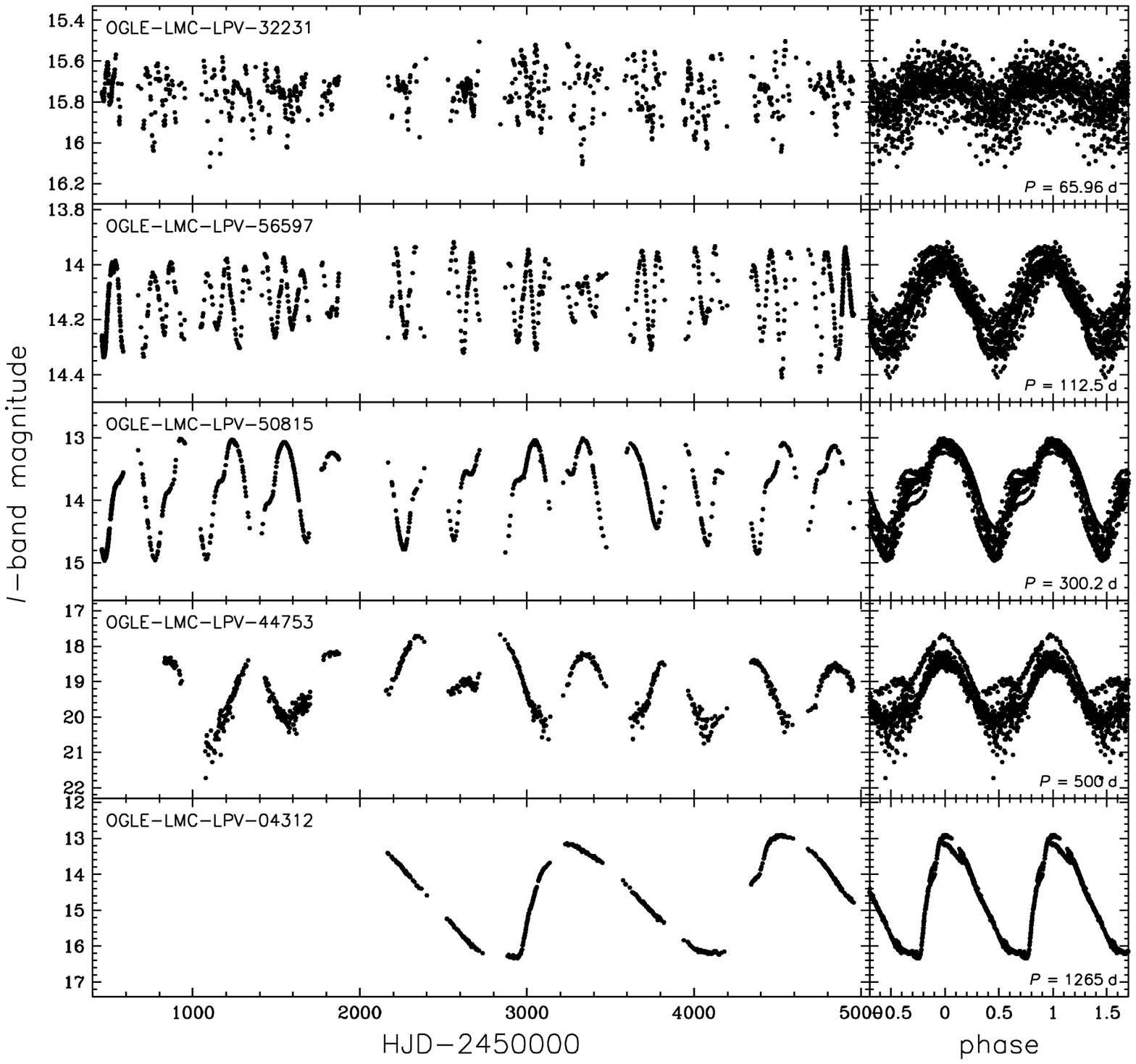}}
\FigCap{Exemplary light curves of stars from PL sequence C. {\it Left panels}
show unfolded OGLE-II (if available) and OGLE-III {\it I}-band light
curves. {\it Right panels} show the same light curves folded with the
primary periods.}
\end{figure}
In total, we detected several dozen candidates for R~CrB stars. Some of
them show characteristic light curves with rapid, deep and irregular
declines in brightness, other candidates are not so evident and demand a
spectroscopic confirmation. However, deeper or shallower declines at
irregular intervals are very common feature in C-rich giants, and this
feature can be helpful in distinguishing between C-rich and O-rich
stars. In this catalog we included only these R~CrB stars which exhibited
any signs of pulsations and it was possible to derive their periods. We
mark these stars in the remarks of the catalog. The full list of candidates
for R~CrB stars will be published in one of the forthcoming parts of the
OIII-CVS.

During the variability selection we detected about 250 very red
($(J-K)>2$~mag) and very faint in the {\it I}-band ($I>18$~mag) Mira
stars. One such light curve (OGLE-LMC-LPV-44753) is shown in Fig.~5. For
the majority of these Mira variables we have no {\it V}-band observations,
because the stars are too faint in this passband. These stars are also
fainter in the $K_s$-band than other Miras with similar periods (Fig.~1),
but most of them lie on sequence~C in the diagram constructed with the
extinction independent $W_{JK}$ index. These objects are heavily obscured
by circumstellar dust shells, like those stars detected by Ita \etal
(2009). Most of these Miras are categorized to C-rich stars, in agreement
with Zijlstra \etal (2006), who noticed that the vast majority of
dust-enshrouded AGB stars belong to this spectral class. Some of these
objects are just above the OGLE detection limit, and there is no doubt that
there must exist stars which are too faint to be detected by our project.

\begin{figure}[b]
\centerline{\includegraphics[width=11.3cm, bb=65 420 535 745]{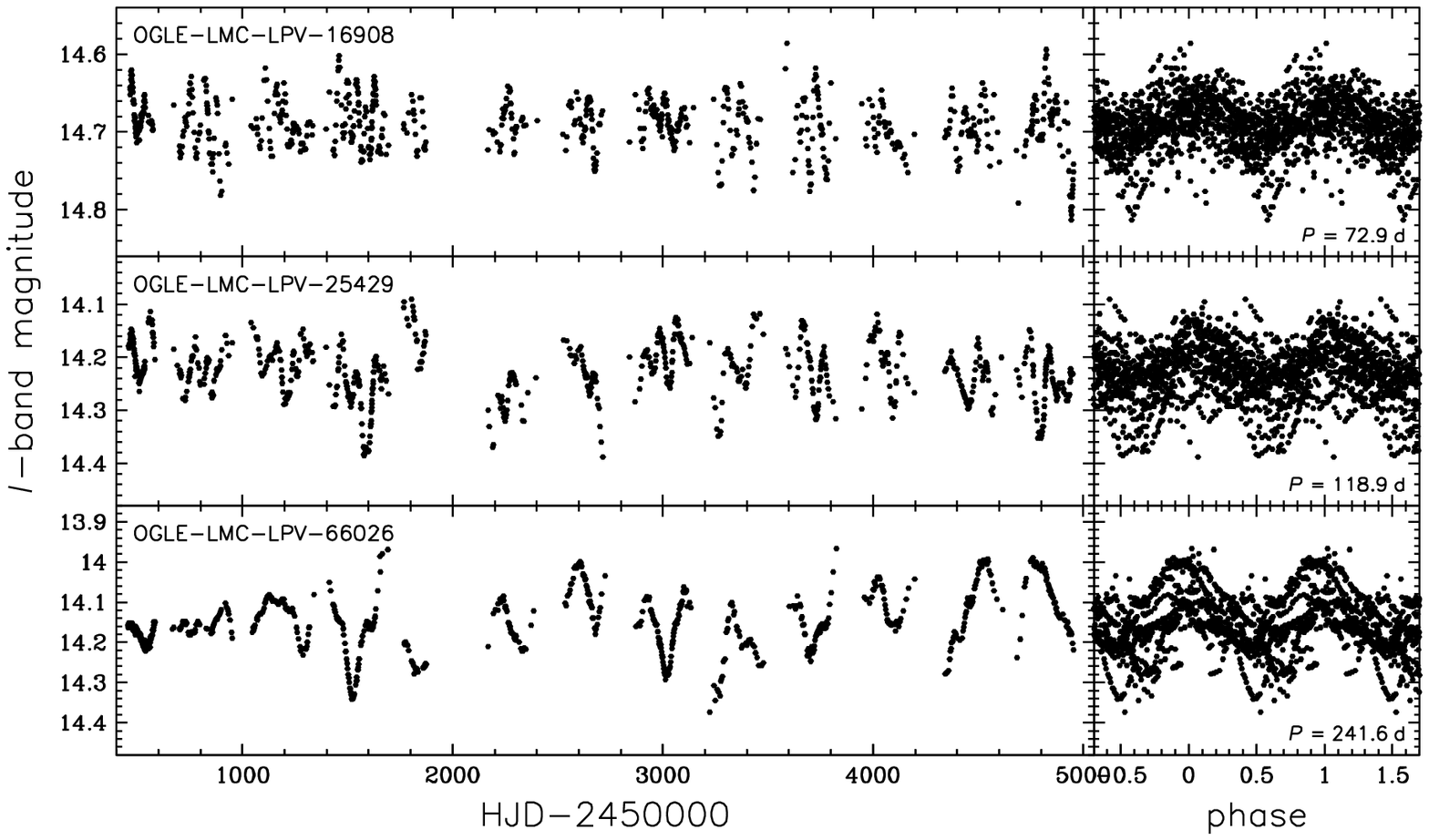}}
\FigCap{Exemplary light curves of stars from the additional PL sequence
located between sequences C and C$'$.}
\end{figure}

In the period--amplitude diagram (Fig.~3) there is a distinct group of 10
Miras with periods longer than 1000~days and with very large amplitudes
($A_I\approx4$~mag). With the exception of one object their light curves
are much more stable than for typical C-rich Miras, so we suppose that
these are O-rich stars. The light curve of one of the longest-period Mira
in our catalog is shown in the bottom panel in Fig.~5. Some of these
objects were discovered by Wood \etal (1992). They suggested that these
long-period Miras are AGB stars with initial masses (on the AGB)
$\apprge4$~\MS.

As it was mentioned in Section 3, it seems that there is an additional PL
sequence between sequences C$'$ and C. It is visible most clearly in the
middle panel ($\log{P}$--$W_{JK}$) of Fig.~1. This dim sequence was first
noticed by Soszyñski \etal (2005). Stars that populate this ridge are SRVs
with generally smaller amplitudes than stars in sequence C. We show three
illustrative light curves in Fig.~6. The secondary periods of these stars
usually fall on sequence C$'$. Further studies are needed to answer
the question, if this additional sequence corresponds to different pulsation
mode than sequence~C, or rather the mode is the same, but stars are somewhat
different. The former
possibility would change the current identification of pulsating modes in
LPVs (stars in sequence C are thought to be fundamental-mode pulsators,
while sequence C$'$ represents the first overtone). The latter hypothesis
means that we have two populations of LPVs in the LMC.

We significantly increased the number of known LPVs in the LMC. The
long-term OGLE photometry is ideal for performing various tests on the
pulsating red giants. We believe that our sample may be used to address the
problem of mode identification in red pulsators, help to resolve the mystery
of long secondary periods in red giants or give clues to the understanding of
the star formation history in the LMC.

\Acknow{
We are grateful to W.~A.~Dziembowski and P.~R.~Wood for helpful comments and suggestions which
improved the paper. We thank R. Gruendl, Z.~Ko³aczkowski, Sz. Koz³owski,
G.~Pojmañski and J.~Skowron for providing the software and data which
enabled us to prepare this study.

This work has been supported by the Foundation for Polish Science through
the Homing (Powroty) Program and by MNiSW grants: NN203293533 to IS and
N20303032/4275 to AU.

The massive period search was performed at the Interdisciplinary Centre for
Mathematical and Computational Modeling of Warsaw University (ICM
UW), pro\-ject no.~G32-3. We are grateful to Dr.~M.~Cytowski for
helping us in this analysis.}

\end{document}